\documentclass{openjournal}
\usepackage{graphicx}
\usepackage{booktabs}
\usepackage{amsmath,amssymb}
\usepackage{orcidlink}
\usepackage{xcolor}
\definecolor{linkcolor}{rgb}{0.0,0.3,0.5}
\usepackage{hyperref}
\newcommand{\zenododoi}{10.5281/zenodo.24628384}
\hypersetup{unicode,colorlinks=true,linkcolor=linkcolor,citecolor=linkcolor,
filecolor=linkcolor,urlcolor=linkcolor,
pdftitle={A long-lived millimetre flare and quiescent variability of the planet-host M dwarf AU Microscopii at 3 mm with ALMA},
pdfauthor={Glenn J. White}}
\graphicspath{{./figures/}}
\journalinfo{Submitted to The Open Journal of Astrophysics}
\submitted{}

\begin{document}
\title{A long-lived millimetre flare and quiescent variability of the
planet-host M dwarf AU\,Microscopii at 3\,mm with ALMA}
\author{Glenn J. White\orcidlink{0000-0002-7126-691X}}
\email{glenn.white@open.ac.uk}
\affiliation{Department of Physics and Astronomy, The Open University,
Walton Hall, Milton Keynes, MK7 6AA, UK}
\affiliation{RAL Space, STFC Rutherford Appleton Laboratory, Chilton,
Didcot, Oxfordshire, OX11 0QX, UK}

\begin{abstract}
We assess how variable the young planet-host M dwarf AU Microscopii is at 3 mm, and ask whether its variability can affect photometry of the debris belt. We use 7.8 h of archival ALMA Band 3 (106 GHz) observations from 2025 January--June. Visibility fits to six quiescent May--June epochs give a mean stellar flux density of 0.28 $\pm$ 0.02 mJy, about twenty times the photospheric prediction; two estimators give intrinsic epoch-to-epoch rms amplitudes of 8 and 11 per cent of the mean, and the belt flux density is 0.94 $\pm$ 0.05 mJy. On January 19 the first 6-s integration of the second of that day's three observations, after a calibration gap, reached 29.2 mJy; the rise was unobserved, and the true maximum may have been higher. The flux fell by a factor of 4.6 in 90 s, but excess emission persisted for about 40 min, with a fitted slow decay time of 30--38 min (depending on fit window and quiescent level) and a recorded isotropic-equivalent Band 3 excess energy of (2.8--3.2) $\times$ 10$^{27}$ erg. On May 26 two peaks 86 s apart reached 10.1 and 10.6 mJy in total flux density, followed by a decay with a fitted slow e-folding time of 8.1 min. Both events are consistent with compact emission at the stellar position within the astrometric uncertainties. The star exceeded our activity threshold during 15.8 per cent of the sampled exposure, mostly on January 19. Flagging the brightest integrations leaves a prolonged stellar excess. With a separate stellar amplitude fitted in each execution, the inferred belt flux differed from its quiescent mean by +5.4 $\pm$ 8.2 per cent in January and $-$3.7 $\pm$ 6.5 per cent in May, including flux-scale uncertainty; neither difference is significant.
\end{abstract}


\section{Introduction}
\label{sec:intro}
AU\,Microscopii (AU\,Mic; M1\,Ve) is a pre-main-sequence M dwarf and
one of the nearest young stars. It lies at $9.714\pm0.002$\,pc
\citep[\textit{Gaia} DR3 parallax $102.94\pm0.02$\,mas;][]{Gaia2023}
in the $\beta$~Pictoris moving group, whose age is $23\pm3$\,Myr
\citep{Mamajek2014}. It is surrounded by an edge-on debris belt,
discovered in scattered light by \citet{Kalas2004} and imaged at
submillimetre and millimetre wavelengths by \citet{MacGregor2013}, \citet{Matthews2015},
\citet{Daley2019} and \citet{Vizgan2022}. Two transiting planets have
been found with the Transiting Exoplanet Survey Satellite
\citep[TESS;][]{Ricker2015,Plavchan2020,Martioli2021}: the
Neptune-sized AU\,Mic\,b and the smaller AU\,Mic\,c, whose
transit-timing variations have been studied by \citet{Szabo2021},
\citet{Wittrock2022} and \citet{Boldog2025}. The star rotates in
4.863\,d \citep[][see also \citealt{Gilbert2022}]{Plavchan2020}. Published radii are
$0.75\pm0.03$\,R$_\odot$ \citep{Plavchan2020} and
$0.82\pm0.02$\,R$_\odot$ \citep{Donati2023}; we carry both through.

Active M dwarfs have gyrosynchrotron coronae that are bright even
between flares, and flares that span many orders of magnitude in energy
and duration \citep{Bastian1998,Gudel2002,Benz2010}. At centimetre
wavelengths AU\,Mic has been followed in detail. \citet{Tristan2025}
found 16 VLA and 3 ATCA flares, from a 30-s burst to a decay lasting
about 5\,h, in a seven-day campaign that also recorded 73 flares from
X-ray to optical wavelengths \citep{Tristan2023}, and \citet{Tristan2026}
split the quiescent 12--25\,GHz spectrum into a steep component
($\alpha=-0.88\pm0.10$, $S_\nu\propto\nu^\alpha$) and a flat
$0.64\pm0.14$\,mJy component. The millimetre record is much thinner.
In archival 2015 ALMA data, \citet{MacGregor2020} detected two short
220-GHz flares from AU\,Mic in the same observing scan. Short bursts have also been seen from Proxima
Centauri \citep{MacGregor2018,MacGregor2021,Howard2022}, for which
\citet{Burton2025} measured flare rates in a 50-h ALMA campaign, and
cosmic-microwave-background surveys catch bright stellar flares at
90--220\,GHz \citep{Guns2021,Naess2021,Tandoi2024,Biermann2025},
though only above tens of mJy. Millimetre flares lasting hours are
known from a few young stars \citep{Bower2003,Massi2006,Salter2008,Salter2010},
and solar
flares have extended millimetre phases \citep{Trottet2002,Luthi2004,Krucker2013}.

Before 2025, AU\,Mic had been observed at millimetre wavelengths only
in debris-disc programmes a few executions long
\citep[e.g.][]{Daley2019,Marino2026}. Twelve ALMA Band~3 executions
taken between 2025 January and June (programme 2024.1.01095.S) are now
public, and give the first extended look at the star at 3\,mm. We use
them to ask three questions. How bright and how steady is the quiet
star? What are its flares like? Does the star affect millimetre
photometry of the belt? The short answers are that the quiet star is
about 20 times brighter than its photosphere, was brighter on May 17
than the 106-GHz extrapolation of a VLA spectrum taken three days
earlier, and changes from day to day; that two large flares were caught, one of them still decaying
40\,min after the brightest recorded integration; and that flagging
the brightest integrations leaves a prolonged stellar excess, while
with a separate stellar amplitude fitted in each execution the
inferred belt flux shows no significant difference from its quiescent
mean.

\section{Observations and data reduction}
\label{sec:obs}

\subsection{The 2025 Band 3 campaign}
\label{sec:campaign}
AU\,Mic was observed with the 12-m array of the Atacama Large
Millimeter/submillimeter Array \citep[ALMA;][]{Wootten2009} in twelve
executions between 2025 January 6 and June 3 (Table~\ref{tab:log}).
Each used three continuum spectral windows centred at 100.9, 102.9
and 112.9\,GHz, giving 5.4\,GHz of usable bandwidth at an effective
frequency of 105.6\,GHz (quoted as 106\,GHz). The four January
executions used a compact configuration (synthesised beams
1.2--2.1\,arcsec) and five target scans, four of 9.4\,min and a final
one of 1.8\,min, recorded as 6.048-s integrations. The eight May--June
executions used an extended configuration (0.7--1.0\,arcsec) and 52--56
target scans of 42\,s, recorded as 2.016-s integrations. Projected
baselines span 13--3600\,m (4.7--1268\,k$\lambda$). The campaign
contains 457 science scans and 7.81\,h of on-source integration. The
observing pattern was fixed in advance, so no scan was triggered by a
flare.

\begin{table*}
\centering
\caption{The twelve Band~3 executions of programme 2024.1.01095.S.
$N_{\rm sc}$ is the number of science scans (102 atmosphere-calibration
scans are excluded by intent); $t_{\rm ret}$ is the retained on-source
integration time (campaign total 7.81\,h). $S_{\rm ref}$ is the
quiescent reference level of the all-baseline light curve
(Section~\ref{sec:ref}): the mean May--June stellar flux plus the belt
flux seen on that execution's own baselines. The below-threshold median is
the median scan flux of the scans below the variability threshold,
with its bootstrap error; for E2 and E3 these scans are flare decay,
not quiescence; $N_{\rm hi}$ counts scans above the
threshold. Full per-scan photometry, including the excess above
$S_{\rm ref}$, is given in an ancillary table.}
\label{tab:log}
\footnotesize
\setlength{\tabcolsep}{4pt}
\begin{tabular}{lcccccccl}
\toprule
Date (2025) & First scan (UT) & Exec. & $N_{\rm sc}$ & $t_{\rm ret}$ (min) & $S_{\rm ref}$ (mJy) & Below-threshold median (mJy) & $N_{\rm hi}$ & Notes\\
\midrule
Jan 06 & 19:00 & E1 & 5 & 39.3 & 0.50 & $0.374\pm0.026$ & 0 & quiescent; star faint\\
Jan 19 & 15:38 & E2 & 5 & 39.3 & 0.44 & $0.84\pm0.08$ & 2 & tail of an earlier flare\\
Jan 19 & 16:36 & E3 & 5 & 39.3 & 0.44 & 0.535 (one scan) & 4 & major flare\\
Jan 19 & 18:28 & E4 & 5 & 39.3 & 0.50 & $0.430\pm0.011$ & 1 & first scan elevated\\
May 17 & 08:09 & E5 & 55 & 38.8 & 0.33 & $0.356\pm0.016$ & 2 & two one-scan excursions\\
May 25 & 10:23 & E6 & 54 & 39.0 & 0.32 & $0.299\pm0.013$ & 1 & one-scan excursion\\
May 26 & 07:52 & E7 & 54 & 39.0 & 0.32 & $0.285\pm0.019$ & 4 & minor episode\\
May 26 & 09:35 & E8 & 55 & 38.8 & 0.32 & $0.322\pm0.016$ & 9 & major flare\\
May 26 & 11:21 & E9 & 53 & 39.2 & 0.32 & $0.328\pm0.025$ & 0 & quiescent\\
May 27 & 10:08 & E10 & 55 & 38.8 & 0.31 & $0.300\pm0.021$ & 0 & quiescent\\
May 28 & 09:10 & E11 & 56 & 38.7 & 0.31 & $0.264\pm0.017$ & 0 & quiescent\\
Jun 03 & 07:38 & E12 & 55 & 38.8 & 0.30 & $0.325\pm0.017$ & 0 & quiescent\\
\bottomrule
\end{tabular}
\end{table*}

\subsection{Calibration and photometry}
\label{sec:cal}
We reduced the raw data in \textsc{casa} 6.7.6 \citep{CASA2022} with
the bandpass, gain and flux calibration of the ALMA quality-assurance
pipeline, the online flags and shadowing flags. The flux scale was set
with J1924$-$2914 in May--June and J2258$-$2758 in January
(J1617$-$5848 for E2). The pipeline adopted calibrator-catalogue models
with spectral indices of $-0.678$ to $-0.714$ (J1924$-$2914), $-0.521$
to $-0.557$ (J2258$-$2758) and $-0.898$ (J1617$-$5848), which also set
the relative scale of the three spectral windows
(Section~\ref{sec:mech}). We take the absolute flux scale to be
uncertain by 5 per cent, mainly through the catalogue flux of the
calibrator. The May--June executions all use J1924$-$2914, so most of
this error is common to them: their six belt fluxes scatter by no more
than their statistical errors, which limits date-to-date changes of
the scale to less than 3.5 per cent (68 per cent confidence). When we
compare the May--June epochs with one another we nevertheless allow an
independent 5 per cent per date, which is conservative. Absolute
fluxes, and comparisons between January (calibrated on J2258$-$2758)
and May--June, carry the full 5 per cent on each side. The phase centre was the \textit{Gaia} DR3 position of
the star propagated to each epoch.

Light curves are visibility-plane averages at the stellar position,
formed per integration and per scan with the delivered weights. All
flux densities are Stokes $I=(XX+YY)/2$. The noise in one integration,
measured from quiescent scans, is 0.41\,mJy at the 6.048-s January
cadence and 0.34\,mJy at the 2.016-s May--June cadence. The data have
no polarisation calibration. The difference of the parallel hands,
$Q'=(XX-YY)/2$, measures one projection of the linear polarisation in
the antenna frame, uncorrected for leakage, so a small $Q'$ does not
exclude strong linear polarisation. The parallel hands carry no
information on circular polarisation, and any linear polarisation of
the calibrators is absorbed by the gain calibration.

\subsection{Archival control data}
\label{sec:controls}
We reduced three archival programmes in the same way: five Band~9
executions of AU\,Mic (2016.1.00878.S; 679\,GHz, 2017 April); three
Band~6 executions of AU\,Mic (2012.1.00198.S; 2014 March and August and
2015 June), each with three 128-channel continuum windows at 213.5,
216.0 and 228.5\,GHz and a 3840-channel window on CO(2--1) at
230.5\,GHz, including the 2015 June flare of \citet{MacGregor2020}; and
one Band~7 execution (2017.1.01583.S; 335\,GHz) of each of three other
$\beta$~Pictoris M-dwarf systems. A 2012 Band~6 execution
(2011.0.00142.S) was not used because we could not recover an absolute
flux scale from its Cycle-0 calibration.

\section{The quiescent star and belt}
\label{sec:quiet}

\subsection{Separating the star from the belt}
\label{sec:starbelt}
The belt contributes to every all-baseline average, by an amount that
depends on the array configuration. We therefore fitted the
visibilities of each execution with a point source at the stellar
position plus a geometrically thin annulus between 9 and 40\,au
\citep{MacGregor2013,Daley2019}, inclined at 88.5\,deg with position
angle 128.48\,deg \citep{Daley2019,Vizgan2022}. The annulus brightens
outward as $I_\nu(r)\propto r^{2.32}$, the profile that
\citet{MacGregor2013} fitted at 1.3\,mm. Our own data prefer a similar
shape: a power law $I_\nu\propto r^\gamma$ fitted to the six stacked
quiescent May--June executions gives $\gamma=2.15$ (68 per cent range
2.0--2.35, with the $\chi^2$ scaled to its minimum reduced value), and
a uniform annulus is excluded ($\Delta\chi^2=227$). We adopt the
published index and quote results for a uniform annulus
where they help to show how much the conclusions depend on the belt
model. The two amplitudes come from a weighted least-squares fit to the
visibilities of the scans below the variability threshold
(Section~\ref{sec:ref}), with jackknife errors over scans (over ten
time blocks for the five-scan January execution), which capture the
scan-to-scan calibration fluctuations that the formal weights miss.
Table~\ref{tab:starbelt} lists the results.

\begin{table}
\centering
\caption{Star and belt flux densities from the visibility fits
(9--40-au annulus with $I_\nu\propto r^{2.32}$, $i=88.5$\,deg,
PA\,$=128.48$\,deg); statistical (jackknife over scans; over ten time
blocks for E1) errors only. $F_\star^{\rm bf}$ refits the star with the
belt fixed at the May--June mean. The six clean May--June executions
define the quiescent star; E7 and E8 contain the minor episode and the
May flare and are shown for comparison. In the compact January
configuration E1 has additional belt-geometry ($\pm0.02$\,mJy) and
belt-profile ($-0.04$\,mJy for a uniform annulus) uncertainties
(Section~\ref{sec:var}).}
\label{tab:starbelt}
\footnotesize
\setlength{\tabcolsep}{3pt}
\begin{tabular}{lcccc}
\toprule
Exec. & Scans & $F_\star$ (mJy) & Belt (mJy) & $F_\star^{\rm bf}$ (mJy)\\
\midrule
E5 & 53 & $0.335\pm0.021$ & $0.938\pm0.037$ & $0.335\pm0.021$\\
E6 & 53 & $0.258\pm0.010$ & $0.915\pm0.035$ & $0.257\pm0.010$\\
E9 & 53 & $0.277\pm0.016$ & $0.968\pm0.042$ & $0.278\pm0.016$\\
E10 & 55 & $0.261\pm0.012$ & $0.922\pm0.043$ & $0.260\pm0.012$\\
E11 & 56 & $0.237\pm0.011$ & $1.005\pm0.042$ & $0.240\pm0.011$\\
E12 & 55 & $0.303\pm0.011$ & $0.911\pm0.062$ & $0.302\pm0.011$\\
\midrule
Mean (six) & --- & $0.279\pm0.014$ & $0.943\pm0.015$ & ---\\
\midrule
E1 (Jan 6) & 5 & $0.181\pm0.024$ & $0.87\pm0.07$ & $0.164\pm0.026$\\
E7 & 50 & $0.273\pm0.015$ & $0.966\pm0.043$ & $0.274\pm0.015$\\
E8 & 46 & $0.310\pm0.020$ & $0.910\pm0.045$ & $0.309\pm0.021$\\
\bottomrule
\end{tabular}
\end{table}

The six quiescent May--June executions give a mean stellar flux density
of $0.279\pm0.014$\,mJy (standard error), with further uncertainties of
$\pm0.014$\,mJy from the absolute flux scale (5 per cent, common to
the six dates), $\pm0.003$\,mJy from the belt geometry and $\pm0.012$\,mJy
from the belt profile, so $F_\star=0.28\pm0.02$\,mJy at 105.6\,GHz. The
geometry term is the spread of the mean over 81 belt geometries (inner
radius 7--11\,au, outer radius 36--44\,au, inclination 88--89\,deg,
position angle $\pm1$\,deg), and the profile term is the change to a
uniform annulus, which gives 0.267\,mJy. A 3700-K Rayleigh--Jeans
photosphere contributes only 12--14\,$\mu$Jy, so the quiet star is
about 20 times brighter than its photosphere.

The belt flux density is $0.943\pm0.015$\,mJy (standard error)
$\pm0.047$\,mJy (flux scale), i.e. $0.94\pm0.05$\,mJy. With the
7.14-mJy belt flux that \citet{MacGregor2013} measured with the same
profile at a mean frequency of 235\,GHz (10 per cent calibration
error), the 1.3--3\,mm spectral index of the belt is
$\alpha_{\rm mm}=2.53\pm0.15$, within the range found for other debris
belts \citep{MacGregor2016}. At 3\,mm the fitted belt does not depend
on the array configuration: the compact January executions E1--E4 give
0.85--0.99\,mJy, against $0.94\pm0.02$\,mJy in the extended
configuration. The index relies on the published calibration of the
2012 data. Our own 2014--2015 Band~6 reductions give belt fluxes of
only 2.2--4.1\,mJy, varying from one execution to the next; their
shortest baselines (9--22\,k$\lambda$) correspond to largest
recoverable scales of 6--14\,arcsec, comparable to the 8-arcsec length
of the belt, which may explain part of the deficit, but we have not
established its cause and do not use them for the belt.

\subsection{Is the quiet star variable?}
\label{sec:var}
The six May--June stellar fluxes range from 0.24 to 0.34\,mJy
(Table~\ref{tab:starbelt}). With their statistical errors and an
independent 5-per-cent flux-scale error on each date, they are
inconsistent with a constant star ($\chi^2=14.0$ for 5 degrees of
freedom, $p=0.016$), while the belt fluxes, which must be constant,
are consistent with one another. A test that does not depend on the
flux scale at all is the ratio of stellar to belt flux within each
execution. Allowing for the anticorrelation of the two fitted
amplitudes, this ratio also rejects a constant, at $p=0.001$--0.003
for either belt profile. Once measurement and flux-scale noise are
removed, two estimators give an intrinsic rms across the six epochs of
0.022 and 0.029\,mJy, or 8 and 11 per cent of the mean. These are two
estimates of one quantity, not a confidence interval. The first is a
maximum-likelihood estimate that treats each date's 5-per-cent flux
scale as an independent error; the second subtracts the mean
statistical variance and a 5-per-cent scale term from the observed
variance. Holding the belt fixed at its mean changes
no stellar flux by more than 0.003\,mJy, so the spread is not a
trade-off between star and belt. Nor is it a pointing effect: fitting
for an offset of the field places the star 0--34\,mas from the phase
centre in the six executions, and allowing for these offsets raises
the stellar fluxes by at most 0.009\,mJy and leaves the test unchanged.
E5 and E10 differ by 0.07\,mJy despite sampling similar rotational
phases (0.00 and 0.07 of the 4.863-d period); the six epochs do not
establish a rotational dependence.

The one quiet January execution, E1 on January 6, gives a fainter
star. Its belt flux, $0.87\pm0.07$\,mJy in the free fit
(Table~\ref{tab:starbelt}), agrees with May--June, but in the compact
configuration the belt seen on E1's baselines is 1.2--2 times the
stellar signal, so the stellar flux depends on the belt model. With the belt fixed at its May--June value,
E1 gives $0.16\pm0.03$\,mJy (statistical) $\pm0.02$\,mJy (belt
geometry), 0.59 of the May--June mean, or 0.12\,mJy for a uniform
annulus. Allowing also for the uncertainty of the May--June mean, the
epoch-to-epoch scatter and an independent 5-per-cent flux scale on
each side, E1 lies 2.4--2.6$\sigma$ below the May--June values
(2.9--3.3$\sigma$ for a uniform annulus). The star was
probably fainter on January 6 than in May and June, but a single
execution cannot establish its quiescent level on January 19.

\subsection{The quiescent centimetre-to-millimetre spectrum}
\label{sec:sed}
\begin{figure}
\centering
\includegraphics[width=0.62\textwidth]{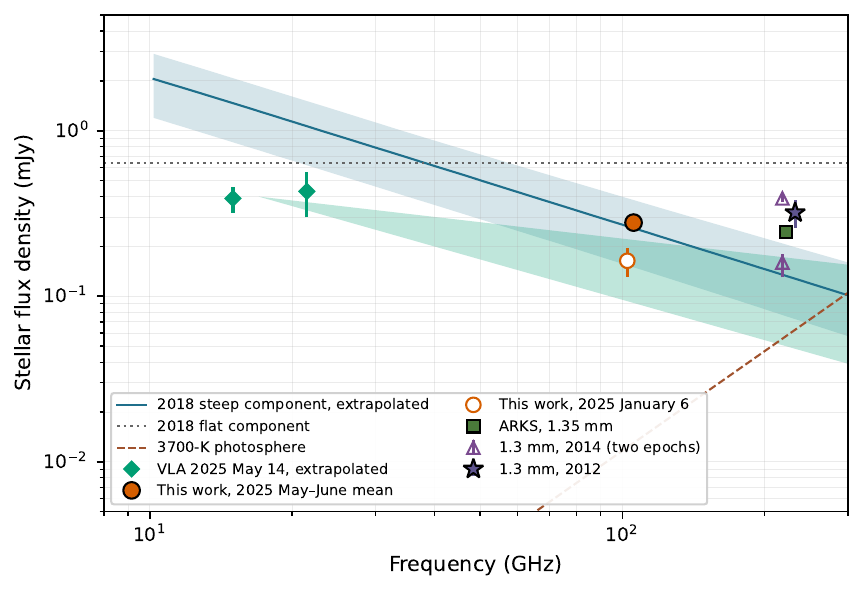}
\caption{Quiescent stellar spectrum of AU\,Mic (non-simultaneous
epochs, 2012--2025). Filled orange circle: the May--June 2025 mean,
$0.28\pm0.02$\,mJy; open circle: 2025 January 6 ($0.16\pm0.03$\,mJy).
Green diamonds and band: the VLA spectrum of 2025 May 14 and its
extrapolation with optically thin indices of $-0.33$ to $-0.81$
\citep{Tristan2026b}. Blue band: the steep 2018 component of
\citet{Tristan2026} (16--84 per cent range); dotted line: their flat
component. Square: ARKS 1.3-mm stellar flux from the combined 2014--2015 data \citep{Marino2026};
triangles: 2014 stellar fluxes \citep{Daley2019}; star: 2012
\citep{MacGregor2013}. Dashed line: 3700-K Rayleigh--Jeans
photosphere.}
\label{fig:sed}
\end{figure}

Fig.~\ref{fig:sed} places the 3-mm star on its quiescent spectrum. The
steep 2018 component of \citet{Tristan2026}, extrapolated from its
15-GHz normalisation (0.56--2.36\,mJy) with $\alpha=-0.88\pm0.10$,
predicts 0.15--0.38\,mJy at 105.6\,GHz (16--84 per cent range of a
Monte Carlo over normalisation and index), too broad a range to test. A
sharper comparison is the VLA 12--26\,GHz spectrum of 2025 May 14,
three days before E5 \citep{Tristan2026b}, which peaks near 17\,GHz at
about 0.4\,mJy and has an optically thin index of $-0.57\pm0.24$.
Extended to 105.6\,GHz with indices of $-0.33$ to $-0.81$, it predicts
0.13 (0.09--0.22)\,mJy. E5 gives $0.335\pm0.021$\,mJy, about 2.6 times
the central prediction. Because the star varies, this one comparison
does not prove a separate millimetre component; we return to the
question in Section~\ref{sec:qdisc}. The 1.3-mm stellar fluxes,
0.16--0.39\,mJy in individual epochs from 2012 to 2015
\citep{MacGregor2013,Daley2019} and $0.245\pm0.014$\,mJy from the
2014--2015 data combined \citep[ARKS;][]{Marino2026}, are similar to the
3-mm flux, but they were taken at other times and cannot fix the
spectral shape. The flat 0.64-mJy centimetre component of
\citet{Tristan2026} would on its own exceed our 3-mm flux, so it must
have faded since 2018 or turn over between 25 and 106\,GHz.

\section{The 2025 flares}
\label{sec:flares}

\subsection{Reference level and variability threshold}
\label{sec:ref}
An all-baseline light curve measures the star plus the part of the belt
seen on that execution's baselines. For each scan we therefore define a
reference level $S_{\rm ref}=F_\star+B\,f_B$, where $F_\star$ and $B$
are the May--June star and belt fluxes and $f_B$ is the fraction of the
belt model recovered on that scan's visibilities (0.02--0.06 in
May--June and 0.16--0.23 in January). $S_{\rm ref}$ is 0.30--0.33\,mJy
in May--June and 0.44--0.50\,mJy in January (Table~\ref{tab:log}), and
the excess above it is the stellar brightening.

\begin{figure*}
\centering
\includegraphics[width=0.85\textwidth]{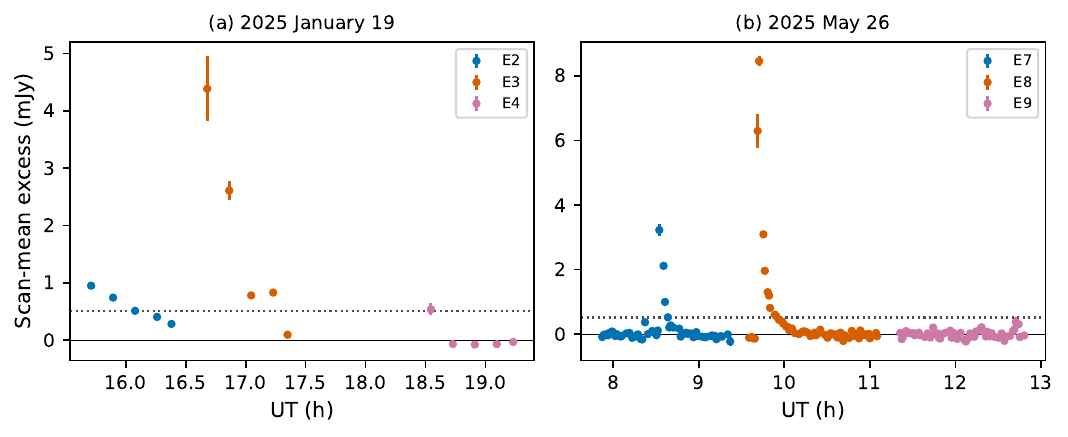}
\caption{Excess above the quiescent reference level $S_{\rm ref}$ for
each scan of the three executions on each flare day, with 1$\sigma$
errors. The dotted line is the variability threshold (0.52\,mJy).
(a) 2025 January 19: the tail of an earlier, unobserved flare (E2), the
major flare (E3) and the following execution (E4). (b) 2025 May 26: the
minor episode (E7), the flare (E8) and a quiescent execution (E9). Scan
means average over up to 9.4\,min (January) or 42\,s (May), so they lie
well below the single-integration peaks of Figs~\ref{fig:jan}
and~\ref{fig:may}.}
\label{fig:lc}
\end{figure*}

Across the 457 science scans the excess has a median of 0.000\,mJy and
a robust scatter (1.4826 times the median absolute deviation) of
0.103\,mJy. We flag a scan as active when its excess exceeds the median
by five times this scatter, i.e. by 0.52\,mJy. The threshold is set
almost entirely by the 437 May--June scans. It is conservative for the
longer January scan means, but a January 42-s interval has a noise of
about 0.16\,mJy, so for the active-time fraction
(Section~\ref{sec:rate}) the threshold is about $3\sigma$ there; no
42-s interval of the quiet January execution E1 exceeds it.
Fig.~\ref{fig:lc} shows the excess light curves of the two flare days.

\subsection{The 2025 January 19 flare}
\label{sec:jan}
\begin{figure}
\centering
\includegraphics[width=0.68\textwidth]{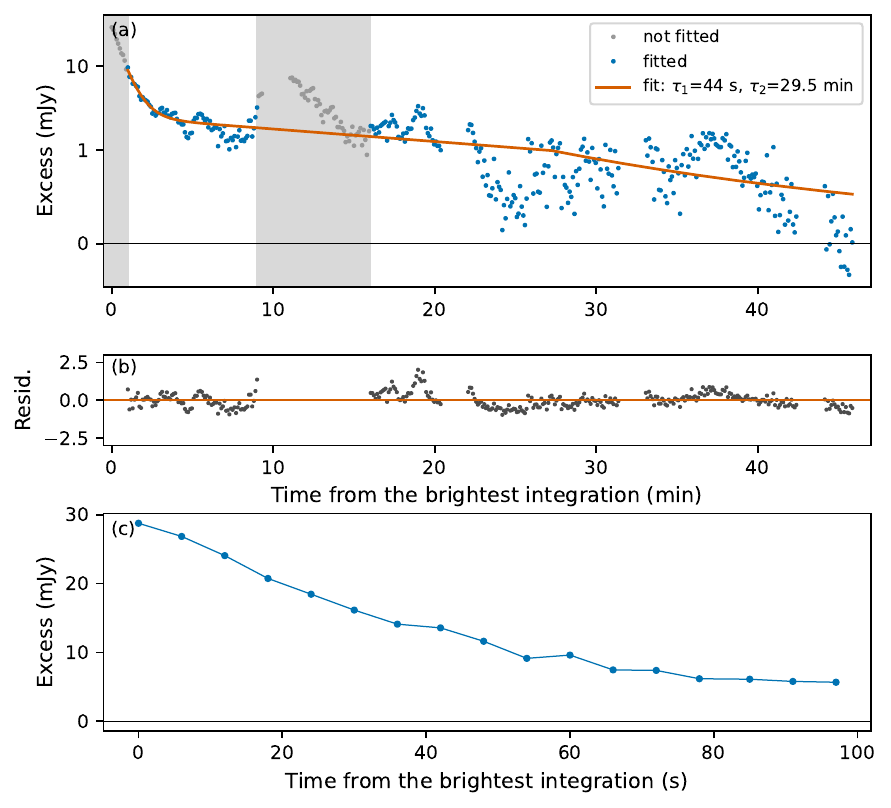}
\caption{The 2025 January 19 flare (E3) at 6.048-s resolution, as
excess above $S_{\rm ref}$. (a) The full execution (logarithmic scale
above 1\,mJy) with the two-exponential fit; grey bands mark the
unfitted first 60\,s and the excluded 9--16-min rebrightening. (b) Fit
residuals. (c) The first 100\,s; the rise before the first integration
was not observed.}
\label{fig:jan}
\end{figure}

Three consecutive executions spanning 3.7\,h caught a flare complex
(Fig.~\ref{fig:lc}a). E2 opens 0.95\,mJy above $S_{\rm ref}$ and falls
through its five scans to 0.28\,mJy, which looks like the tail of an
earlier flare. The next 12\,min contain only calibration scans, so the
onset of the main flare was missed. The first integration of E3,
centred at 16:36:35\,UT, is the brightest recorded:
$29.2\pm0.4$\,mJy (statistical) $\pm1.5$\,mJy (calibration), about 105
times the quiet star (Fig.~\ref{fig:jan}c). The true maximum may have been
higher. The flux fell to 14\,mJy within 40\,s and to 6.4\,mJy by 90\,s,
a factor of 4.6, and then decayed slowly for the rest of the execution
(Fig.~\ref{fig:jan}a).

We fitted the excess with two exponentials,
$S(t)=S_{\rm ref}+A_1e^{-t/\tau_1}+A_2e^{-t/\tau_2}$, with $t$
measured from the brightest integration, from 60\,s to the end of the
execution at 2751\,s. We left out 9--16\,min, which contains a separate
rebrightening to 7.7\,mJy at 11\,min, about 5\,mJy above the smooth
decay. Because the residuals flicker and stay correlated for several
minutes, we include an extra noise term and take uncertainties from a
bootstrap over 255-s blocks (42-s blocks for $\tau_1$). The fit gives
$\tau_1=44$ [37, 51]\,s and $\tau_2=29.5$ [22.8, 35.7]\,min (68 per
cent ranges; amplitudes in Table~\ref{tab:flares}). At 60\,s it
predicts 9.3\,mJy against the measured 10.0\,mJy.

The slow decay time depends on two choices. If the January star was as
faint as on January 6, the lower reference level gives $\tau_2=32.3$ or
33.2\,min. Ending the fit at 2400\,s gives 33.4--37.6\,min, because the
last 6\,min of the execution fall below the extrapolated curve. The
slow decay time is therefore 30--38\,min. The residuals
(Fig.~\ref{fig:jan}b) show flickering of $\pm$0.5--1\,mJy on
time-scales of minutes, so the decay is not purely exponential. The key
result is that emission well above the quiet level lasted at least
40\,min after a fast initial decline that would be called impulsive at
shorter wavelengths.

The next execution, E4, began 1\,h\,52\,min after the brightest
integration of E3. Its first scan is 0.54\,mJy above $S_{\rm ref}$,
above the threshold and 5.7$\sigma$ above the median of its later
scans, after which the star is quiet. The slow component of the E3
fit predicts only 0.06\,mJy by then, but with the intervening gap we
cannot tell whether E4 saw the end of the same event.

\subsection{The 2025 May 26 flare}
\label{sec:may}
\begin{figure}
\centering
\includegraphics[width=0.68\textwidth]{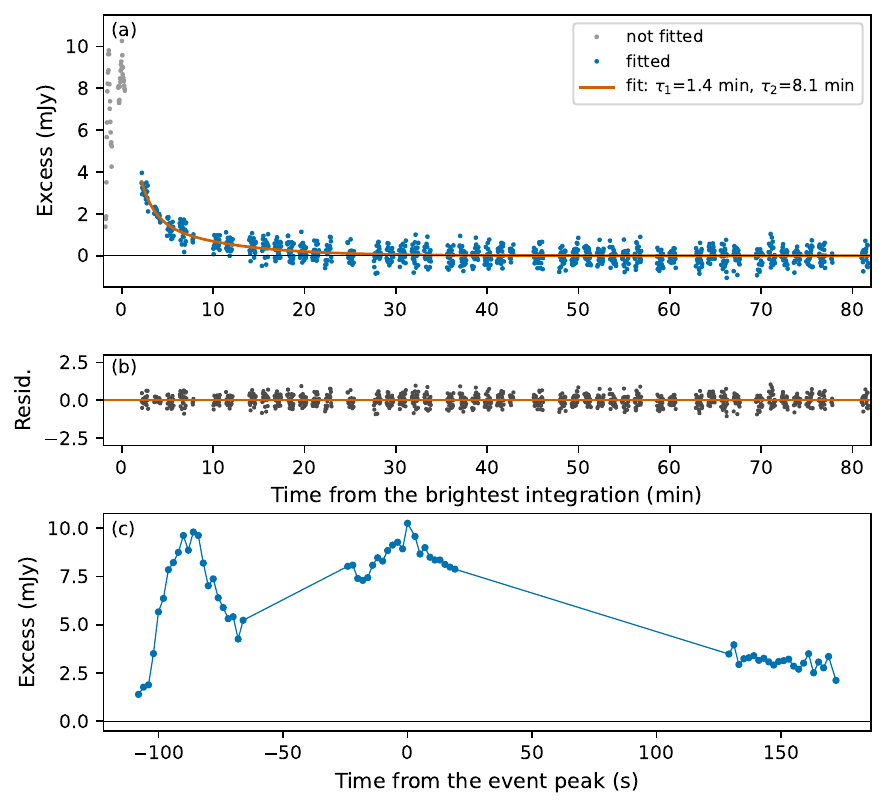}
\caption{The 2025 May 26 flare (E8) at 2.016-s resolution, as excess
above $S_{\rm ref}$. (a) The full execution with the two-exponential
fit to the integrations after the peak. (b) Fit residuals. (c) The two
peaks, 86\,s apart; gaps are calibration scans.}
\label{fig:may}
\end{figure}

The May flare (E8, 09:35:36--11:05:31\,UT), recorded at 2.016-s
cadence, shows two peaks 86\,s apart (Figs~\ref{fig:lc}b
and~\ref{fig:may}c). The star was quiet in the last integration of the
scan ending at 09:40:32, and burst~1 was already rising when the next
scan began at 09:41:15. Its rise is resolved: the excess grows from
about 1.5\,mJy to 8--10\,mJy within 12\,s, and peaks at 9.8\,mJy
(10.1\,mJy total). The second scan opens already at about 8\,mJy, so
the onset of burst~2, and whether the flux fell back between the
peaks, were not observed. Burst~2 peaks at 09:43:03 at
$10.6\pm0.3$\,mJy (statistical) $\pm0.5$\,mJy (calibration), 38 times
the quiet star. After a 110-s calibration gap the next science scan
opens at 3.8\,mJy.

The same two-exponential model, fitted from the first integration
after the gap (129\,s after the peak) to the end of the execution,
gives $\tau_1=1.4$ [0.9, 1.6]\,min and $\tau_2=8.1$ [7.3, 9.0]\,min. The
May residuals are uncorrelated, so 42-s blocks suffice for the
bootstrap.
At 129\,s the model predicts 3.85\,mJy against 3.80\,mJy measured.
Because no data exist between 19 and 129\,s after the peak, the fast
component is poorly constrained, but the 8-min slow decay is well determined.
The May decay is therefore slower than the bursts but about four times
faster than the January tail: the fitted excess falls below 0.3\,mJy
17\,min after the peak, and E9, starting 1\,h\,38\,min after the peak,
is quiet. A four-scan minor episode in the preceding execution (E7)
reached an excess of 3.2\,mJy in its brightest scan (5.3\,mJy total in
one integration), 68\,min before the flare.

\begin{table}
\centering
\caption{Properties of the two flares. Peak flux densities are the
total (star, belt and flare) all-baseline flux of the brightest single
integration, with statistical and calibration errors; $L_\nu$ and
$T_{\rm b}$ use them, which overstates the flare alone by 1--3 per cent. Brightness temperatures assume emission filling the stellar disc
(0.75--0.82\,R$_\odot$). Decay parameters are from the
two-exponential fits (68 per cent block-bootstrap ranges). In-band
indices $\alpha$ are fitted to the excess flux in the three spectral
windows (Section~\ref{sec:mech}); errors are statistical, and an
assumed calibration sensitivity of $\pm0.1$ per epoch applies in
addition. $^a$32.3--33.2\,min with the January 6 stellar level as
baseline, and 33.4--37.6\,min with the fit ending at 2400\,s. $^b$January: first five integrations; May: through both bursts, consistent with noise. $^c$January decay, rebrightening excluded: the
time mask of the decay fit (60--2751\,s after the brightest integration,
omitting 9--16\,min). $^d$Of which $3.7\times10^{26}$\,erg is a
backward extrapolation of the decay fit across the 108-s gap after the
brightest integration.
Energies are defined in Section~\ref{sec:energy}.}
\label{tab:flares}
\footnotesize
\setlength{\tabcolsep}{3pt}
\begin{tabular}{lcc}
\toprule
 & 2025 Jan 19 (E3) & 2025 May 26 (E8)\\
\midrule
Peak $S_\nu$ (mJy) & $29.2\pm0.4\pm1.5$ & $10.6\pm0.3\pm0.5$\\
Integration (s) & 6.048 & 2.016\\
$L_\nu$ at peak (erg\,s$^{-1}$\,Hz$^{-1}$) & $3.3\times10^{15}$ & $1.2\times10^{15}$\\
$T_{\rm b}$ (disc-filling, K) & (7.5--9.0)$\times10^6$ & (2.7--3.3)$\times10^6$\\
$A_1$ (mJy), $\tau_1$ & 25.2, 44 [37, 51]\,s & 8.2, 1.4 [0.9, 1.6]\,min\\
$A_2$ (mJy), $\tau_2$ & 2.51, 29.5 [22.8, 35.7]\,min$^a$ & 2.34, 8.1 [7.3, 9.0]\,min\\
$\alpha$ at peak (Jan 0--18\,s; May $\pm20$\,s) & $-0.40\pm0.18$ & $-1.25\pm0.25$\\
$\alpha$ in decay$^c$ & $-0.96\pm0.29$ & $-1.8\pm0.6$ (2--10\,min)\\
Max.\ measured $|Q'|/I$ (one projection)$^b$ & 0.013 & 0.03--0.04\\
$E_{\rm rec}$ (erg) & (2.8--3.2)$\times10^{27}$ & $6.8\times10^{26}$\\
$E_{\rm rec+gaps}$ (erg) & $3.1\times10^{27}$ & $1.4\times10^{27}$$^d$\\
$E_{\rm model}$ (erg) & $3.4\times10^{27}$ & $1.1\times10^{27}$\\
$E_{\rm rec}/$(quiet-star output) & 7.4 & 1.7\\
\bottomrule
\end{tabular}
\end{table}

\subsection{Energies}
\label{sec:energy}
The isotropic-equivalent radiated energy in Band~3 is
\begin{equation}
E=4\pi d^2\,\Delta\nu\sum_i (S_i-S_{{\rm ref},i})\,\Delta t_i ,
\label{eq:E}
\end{equation}
summed over integrations, with $\Delta\nu=5.4$\,GHz and $d=9.714$\,pc,
so that 1\,mJy for 1\,s corresponds to $6.10\times10^{23}$\,erg. The
recorded energy, $E_{\rm rec}$, sums every integration of the
execution: E3 from its first integration to 2751\,s, and E8 from
447\,s before the peak (including burst~1) to 4948\,s after it. Three
short E8 scans (four integrations each; Section~\ref{sec:rate}) have
no stored per-integration light curve, and each enters once, at its
scan-mean excess times its 8.06-s retained time.

For January the result depends on the reference level:
$2.97\times10^{27}$\,erg above $S_{\rm ref}$, $2.83\times10^{27}$\,erg
above the level of the last E3 scan (still in the decay) and
(3.14--3.20)$\times10^{27}$\,erg above the January 6 levels, with a
further 5-per-cent flux-scale uncertainty. Of the energy recorded above
$S_{\rm ref}$, 27 per cent arrived in integrations centred within the
first 90\,s and 73 per cent
afterwards, including 22 per cent of the total in the 9--16-min
rebrightening. E3 spans 46.0\,min, of which 39.3\,min was observed;
the rest is four calibration gaps of 96--100\,s. Filling the three gaps
after the rebrightening with the fitted decay adds
$0.16\times10^{27}$\,erg, giving $E_{\rm rec+gaps}=3.1\times10^{27}$\,erg;
the gap inside the rebrightening is left unfilled. The fitted decay
integrated from the brightest integration to infinity gives
$E_{\rm model}=3.4\times10^{27}$\,erg. E4 adds a further
(1.1--2.1)$\times10^{26}$\,erg, depending on whether its reference is
$S_{\rm ref}$ or the median of its later scans.

In May the recorded excess is $6.77\times10^{26}$\,erg for either
reference level: $1.70\times10^{26}$\,erg in the burst-1 scan,
$2.29\times10^{26}$\,erg in the burst-2 scan, $2.90\times10^{26}$\,erg
in the decay from 129\,s after the peak and $-0.13\times10^{26}$\,erg in
the earlier quiet scans. The three scans without stored integrations
add a net $0.03\times10^{26}$\,erg, all in the decay; one of them,
7\,min after the peak, has an excess of 0.82\,mJy, close to the
0.87\,mJy of the fitted decay at that time. The two bursts carry 59
per cent of what was recorded. E8 spans 90.0\,min, of which 38.8\,min
was observed. Filling the calibration gaps after the peak with the
fitted decay, excluding the intervals covered by those scans, adds
$6.9\times10^{26}$\,erg ($E_{\rm rec+gaps}=1.4\times10^{27}$\,erg).
More than half of this, $3.7\times10^{26}$\,erg, falls in the 108-s
gap that begins 20\,s after the brightest integration, before the
first data used in the decay fit (129\,s); it is a backward
extrapolation of the fit across the unobserved early decay, and
$E_{\rm rec+gaps}$ should be read with that in mind. The five gaps
before the peak, including the 40\,s between the two burst scans, are
left unfilled. The fitted decay alone, from the peak to infinity, gives
$E_{\rm model}=1.1\times10^{27}$\,erg; it excludes burst~1 and relies on
the same extrapolation. We therefore take $E_{\rm rec}$ as the May
result. For
comparison, the quiet star radiates $4.0\times10^{26}$\,erg in the
band over one execution, so the January flare emitted 7.4 times, and
the May flare 1.7 times, the quiescent output.

\subsection{Location and checks}
\label{sec:valid}
Both events are consistent with compact emission from AU\,Mic. To
locate them we subtracted the quiescent star-plus-belt model and fitted
a point source of free position to the visibilities of the bright
integrations within 60\,s of the brightest one (for May, the burst-2
scan; the 1--45-min decay was fitted separately in January). Our visibility extracts hold only the real part of the
phase-centred visibilities, which fixes an offset only up to its sign
and cannot tell a single offset point from a symmetric pair of points.
For May burst~2, the fitted offset is 0.012\,arcsec, consistent with
zero within the 0.034-arcsec 68 per cent radius; the E7 episode lies
at 0.03\,arcsec (68 per cent radius 0.06\,arcsec). In January the peak
lies $0.10\pm0.02$\,arcsec from the phase centre and the decay
(1--45\,min) $0.08\pm0.04$\,arcsec, along similar position angles
modulo 180$^\circ$ (the real visibilities respond identically to
offsets $+\boldsymbol{\theta}$ and $-\boldsymbol{\theta}$); the
two phases of the flare, which share the phase-referencing errors of
one execution, therefore agree. These are formal fitting errors; the
absolute astrometric accuracy of phase referencing in the compact
January configuration is of order a tenth of the 1.2--2.1-arcsec beam.
The 40-s decline gives a light-crossing scale of at most 0.008\,arcsec
for the rapidly varying component. The fitted January displacement is
comparable to the astrometric uncertainty, and 0.10\,arcsec is 0.97\,au
at the star, 15 times the orbit of AU\,Mic\,b, so a real displacement
of a stellar flare by this amount is implausible. The January data are
therefore consistent with a compact flare at AU\,Mic; they do not
establish a physical displacement. If the whole field is offset, the
January all-baseline fluxes are 2 per cent low, well within the
flux-scale uncertainty, and we do not correct the light curves for it.
The fitted offsets are listed in \texttt{table\_s6\_positions.csv}.

\begin{figure}
\centering
\includegraphics[width=0.85\textwidth]{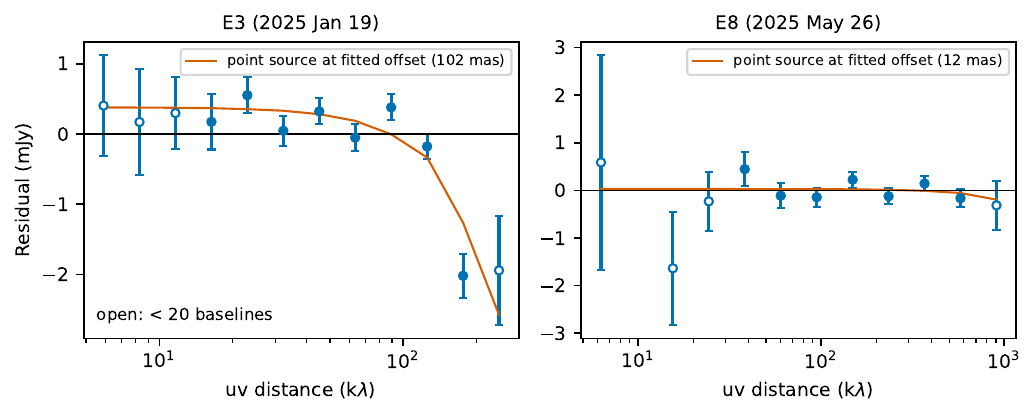}
\caption{Residual real part of the visibility against baseline length
for the flare peaks ($\pm60$\,s of the brightest integration), after
subtracting the quiescent star-plus-belt model and each integration's
own mean, in logarithmic bins. Filled points have at least 20
baselines per integration and spectral window; open points have fewer.
Left: January (E3); right: May (E8), burst~2 (burst~1 lies outside the
window). The orange line is the residual
expected for a point source at the fitted position (0.10\,arcsec from
the phase centre in January, 0.012\,arcsec for May burst~2). With real
visibilities alone, an offset point source, a symmetric source of
similar angular size or residual phase decorrelation produce similar
deficits.}
\label{fig:uv}
\end{figure}

To look for spatial structure we subtracted the quiescent model and
each integration's own mean and binned the residual visibilities by
baseline length, using the integrations within 60\,s of the brightest
one (Fig.~\ref{fig:uv}); in May this window contains burst~2 only. There
the residuals are
consistent with zero from 5 to 1136\,k$\lambda$ ($\chi^2=8.8$ for 11
bins), and every well-sampled bin lies within 0.45\,mJy (4 per cent of
the peak) of zero, so burst~2 is unresolved even on the
longest baselines. In January the two bins beyond 150\,k$\lambda$ lie
about 2\,mJy below zero. A point source at the fitted 0.10-arcsec
offset reproduces this deficit ($\chi^2$ falls from 61 to 16 for 12
bins), although with real visibilities alone a symmetric source of
similar angular size, or residual phase decorrelation on the longest
baselines, would produce a similar deficit.

Several checks rule out instrumental causes. Stacking the first
integration of every quiescent scan gives $0.278\pm0.015$\,mJy, so
there is no artefact at the start of scans. Phase-only
self-calibration of the flare scans changes the peaks by less than 0.5
per cent. No antenna pair is unusually noisy, and dropping the noisiest
quarter changes the May peak by less than 0.5 per cent. The phase
calibrators are stable to about 1 per cent through each execution. The
two parallel hands agree at both events: over the first five January
integrations the largest measured $|Q'|/I$ is 0.013. In May both hands
rise together through the resolved rise of burst~1; $|Q'|/I$ is
0.03--0.04 at the two burst maxima, and single integrations through
both bursts scatter within $\pm0.06$, as expected from the noise.

\section{Flare occurrence and the archival record}
\label{sec:occ}

\subsection{How often does AU Mic flare at 3 mm?}
\label{sec:rate}
Dividing the excess light curves of each scan into consecutive 42-s
intervals from the start of the scan, with a final partial interval
counted at its own duration,
the star was above the threshold for 15.8 per cent of the sampled
exposure (73.9 of 468.5\,min, the retained on-source time of
Table~\ref{tab:log}), but for only 3.3 per cent (11.7 of 350.5\,min)
outside January 19. Three short E8 scans of four integrations each
(24.2\,s in total) have no stored per-integration light curve; each
enters as a single interval at its scan-mean flux, and one of them
(8.1\,s, during the May flare) is above the threshold. The January 19 executions contribute
62.2\,min: 24.3\,min in E2, 35.1\,min in E3 and 2.8\,min in E4. Because
the intervals lie within scans, a scan whose mean is below the
threshold can still contribute. This fraction describes our sample,
not a long-term duty cycle. In 37 of the 457 science scans the star
was at least twice its quiescent brightness.

Two events with recorded integrations above 10\,mJy were seen in
7.81\,h on source. We do not convert this into a flare rate: the 10-mJy
level was chosen after looking at the data, the January flare was
already bright when the observations resumed, and the effective time
over which an onset could have been caught is not defined for events
that last tens of minutes. Besides the two flares, the threshold picked out three
minor episodes and one candidate: four scans in E7 with excesses of
0.52--3.22\,mJy (10--21 times each scan's standard error above that
execution's quiet level), single scans in E5 with excesses of 0.68 and
0.74\,mJy (6.5 and 6.1 times their errors), and a single E6 scan with
an excess of 0.90\,mJy (3.9 times its error). Allowing for the 457 scans
searched, the E6 excursion has $p\simeq0.02$ under our noise model, and
we treat it as a candidate.

We tested whether a short spike like the 2015 flare would have been
picked out by the scan-level threshold. We added a Gaussian of 35\,s full width at half maximum
and 17 or 23\,mJy peak (the 2015 peak scaled from 220 to 106\,GHz with
indices of 0 to $-0.4$) to the integration light curves, centred every
10\,s through each execution, and counted a recovery when it lifted a
quiet scan above the threshold. Injections into scans that were
already active count as misses. Of the injections centred inside a
science scan, 1417 of 1431 were recovered in the six quiet May--June
executions (all the misses fall in the three scans already active),
197 of 236 in the flare execution E8, and 235--237 of 237 in E1, whose
long scans dilute the spike but still exceed the threshold. Counting
centres across the whole elapsed time, including calibration gaps,
85--86 per cent were recovered in the quiet May--June executions, 71--72
per cent in E8 and 86--87 per cent in E1. These figures apply to spikes
of this shape and size. No other spike of comparable recorded amplitude
is present in the data.

The recorded 3-mm energies, (2.8--3.2)$\times10^{27}$ and
$6.8\times10^{26}$\,erg in 5.4\,GHz, correspond to 4.1--4.7 and
$1.0\times10^{27}$\,erg in the 8-GHz bandwidth convention of
\citet{Burton2025}, if the spectrum is flat across the band. The
January flare is above the top of the $10^{24}$--$10^{27}$\,erg range of
the 463 Proxima Centauri flares that \citet{Burton2025} detected at
233\,GHz in 50\,h, and the May flare is near its upper end, although
the comparison is between different stars and frequencies.

\subsection{Archival controls}
\label{sec:arch}
We detected no comparable long-lived excess in the archival controls
(Table~\ref{tab:arch}). In the 2017 Band~9 data no scan exceeds its
execution mean by $5\sigma$, but the scan sensitivity of 5--10\,mJy
would have missed a January-like event diluted over a scan. In the
2014 Band~6 data the largest deviation is the end of a steady 25-min
rise in the March execution, more likely slow stellar variability or
gain drift than a flare.

Our reduction recovers the primary 2015 June flare \citep{MacGregor2020} at
52$\sigma$ in its scan. Its peak in our 6.048-s integrations is
13.3\,mJy, with the two hands agreeing (13.7 and 13.0\,mJy), against the
published $16.8\pm0.4$\,mJy from point-source fits at about 5-s
resolution. We use the same three 2-GHz continuum windows (adding the
narrow CO window changes the peak by 0.2 per cent). Repeating the
published method, a point-source fit to baselines longer than
100\,k$\lambda$, gives 12.0\,mJy at our 6.048-s sampling (the
published value came from 5-s intervals), 29 per cent below the
published value. We cannot explain this difference, and it limits any
comparison of the 2015 and 2025 peak fluxes. A constant amplitude error
of this kind in our own extraction would leave the 2025 decay times
unchanged but would scale the 2025 peak luminosities and energies; the
2025 flux scale is tied to the calibrator models of Section~\ref{sec:cal}.

The 2015 data would have shown a January-type tail. The slow January
component 3--5.4\,min after the peak (2.2\,mJy at 106\,GHz), scaled to
220\,GHz with the January decay index ($-0.96\pm0.29\pm0.1$), predicts
0.9--1.4\,mJy, whereas the 23 integrations 3--5.4\,min after the 2015
peak average $0.44\pm0.07$\,mJy, the same as the level before the
flare. A tail as bright as the scaled January one was therefore absent
in 2015 over that interval; weaker or later emission is not
constrained.

The Band~7 data of the three comparison systems reproduce the published
fluxes of AT\,Mic\,A and B, GJ\,2006\,A and the GSC\,07396$-$00759 disc
\citep{Cronin2022,Cronin2023} to within $1.1\sigma$, which validates the
reduction, and show no flares. Their $5\sigma$ limits per scan are
0.26\,mJy for each AT\,Mic component, 0.28\,mJy for GJ\,2006 and
0.70\,mJy for GSC\,07396$-$00759. A January-type flare moved to these
distances and to 335\,GHz would peak in a 6-s integration at 18, 1.5
and 0.35\,mJy. Averaged over a scan as long as ours, which diluted the
January peak about sixfold, only the AT\,Mic limit would catch it.

\begin{table}
\centering
\caption{Archival AU\,Mic control executions, reduced as for the 2025
data: scan means (including the belt), weighted by the inverse square
of each scan's standard error, and the range of scan means. For 2015
June 24 the mean and range are for the four quiescent scans; the flare
scan, with a mean of 2.39\,mJy, is listed separately.}
\label{tab:arch}
\footnotesize
\begin{tabular}{lccc}
\toprule
Date (UT) & $N_{\rm sc}$ & Mean (mJy) & Range (mJy)\\
\midrule
\multicolumn{4}{l}{Band 9 (679\,GHz), 2016.1.00878.S}\\
2017 Apr 22 08:54 & 12 & $1.34\pm0.24$ & $-0.26$--3.03\\
2017 Apr 23 09:26 & 11 & $2.17\pm0.32$ & 0.90--4.25\\
2017 Apr 23 11:27 & 11 & $1.37\pm0.29$ & $-3.15$--4.40\\
2017 Apr 26 08:52 & 11 & $1.41\pm0.27$ & $-1.60$--3.16\\
2017 Apr 26 12:09 & 9 & $1.76\pm0.29$ & $-0.44$--5.69\\
\multicolumn{4}{l}{Band 6 (213.5--230.5\,GHz), 2012.1.00198.S}\\
2014 Mar 26 09:58 & 5 & $0.76\pm0.02$ & 0.67--0.91\\
2014 Aug 18 05:40 & 5 & $0.24\pm0.03$ & 0.10--0.29\\
2015 Jun 24 04:07 & 4 & $0.45\pm0.02$ & 0.41--0.50\\
\quad flare scan & 1 & 2.39 & ---\\
\bottomrule
\end{tabular}
\end{table}

\subsection{Comparison with other millimetre flares}
\label{sec:compare}
Table~\ref{tab:compare} compares the AU\,Mic events with selected
published Proxima Centauri flares. The January peak spectral luminosity,
$3.3\times10^{15}$\,erg\,s$^{-1}$\,Hz$^{-1}$ at 106\,GHz, is 1.7--2.4
times that of the 2015 AU\,Mic flare at 220\,GHz (using the published
and our re-measured 2015 peaks) and 14 times the brightest Proxima
flare at 233\,GHz \citep{Burton2025}; untargeted surveys reach higher
luminosities in rarer, more distant events
\citep{Guns2021,Naess2021,Tandoi2024}. Scaling the mean of the first
35\,s of the January flare (22.9\,mJy) to 220\,GHz with the peak index
gives 17 (15--20)\,mJy, comparable to the published 2015 peak, but the
January rise was not observed, so the two impulsive phases cannot be
compared directly. Long decays are well known at centimetre
wavelengths, where \citet{Tristan2025} saw an AU\,Mic flare decay over
about 5\,h, and at 90\,GHz in the young binary V773\,Tau\,A, whose
flare decayed with an e-folding time of 2.3\,h \citep{Massi2006}.

\begin{table*}
\centering
\caption{The AU\,Mic millimetre flares compared with selected published
Proxima Centauri flares. Spectral
luminosities are as published or computed at the same distances.
Characteristic times ($t_{1/2}$, full width at half maximum): the 2015
AU\,Mic events last $t_{1/2}\simeq35$ and 9\,s; the Proxima events listed here $t_{1/2}=4.3$--16\,s; the 2025 events
decay with e-folding times of 44\,s and 29\,min (January) and 1.4 and
8.1\,min (May). For the 2015 primary flare our reduction gives a peak of
12.0\,mJy (point-source fit to baselines longer than 100\,k$\lambda$;
Section~\ref{sec:arch}).}
\label{tab:compare}
\footnotesize
\begin{tabular}{lcccl}
\toprule
Event & $\nu$ (GHz) & Peak flux (mJy) & $L_\nu$ (erg\,s$^{-1}$\,Hz$^{-1}$) & Reference\\
\midrule
AU Mic 2015 Jun 24 (primary) & 220 & $16.8\pm0.4$ & $1.9\times10^{15}$ & \citet{MacGregor2020}\\
AU Mic 2015 Jun 24 (secondary) & 220 & $6.07\pm0.35$ & $6.9\times10^{14}$ & \citet{MacGregor2020}\\
Proxima 2019 May 2 & 233 & 47 & $9\times10^{13}$ & \citet{Burton2025}\\
Proxima 2019 May 1 & 225--241 & $106\pm8$ & $2.1\times10^{14}$ & \citet{MacGregor2021}\\
Proxima 2017 Mar 24 & 233 & $119\pm7$ & $2.4\times10^{14}$ & \citet{MacGregor2018,Burton2025}\\
AU Mic 2025 Jan 19 & 106 & $29.2\pm0.4\pm1.5$ & $3.3\times10^{15}$ & this work\\
AU Mic 2025 May 26 & 106 & $10.6\pm0.3\pm0.5$ & $1.2\times10^{15}$ & this work\\
\bottomrule
\end{tabular}
\end{table*}

\section{Discussion}
\label{sec:disc}

\subsection{Emission mechanism}
\label{sec:mech}
If the flaring plasma covered the stellar disc, the peak brightness
temperatures would be (7.5--9.0)$\times10^6$\,K in January and
(2.7--3.3)$\times10^6$\,K in May (Table~\ref{tab:flares}). These are
lower limits for any source no larger than the disc, and reach about
$10^9$\,K for a loop of $0.1R_\star$. The resolved 12-s May rise limits
the source to $<3.6\times10^{11}$\,cm, about seven stellar radii, which
is not constraining. The emission is therefore compatible with
incoherent gyrosynchrotron radiation from mildly relativistic electrons
\citep{Dulk1985,Bastian1998,Gudel2002}, but without full polarisation
or simultaneous optical, ultraviolet or X-ray data this is not proven.
Emission at 106\,GHz needs gyrofrequency harmonics of order 10--20 in
kilogauss fields, consistent with the 2.1--2.3-kG mean field
\citep{Kochukhov2020} and $\sim$4-kG small-scale fields
\citep{Donati2025} of AU\,Mic, or MeV electrons in weaker fields
\citep{MacGregor2020}.

The spectral indices are measured across the three spectral windows
(100.9, 102.9 and 112.9\,GHz) from identical integrations. Each
window's quiescent reference is the May--June star, assumed flat
across the band, plus the belt seen on that window's own visibilities,
scaled with the belt index; the window fluxes are given in an
ancillary table. Both flares have negative indices: $-0.40\pm0.18$ at
the January peak (the four integrations 0--18\,s after the brightest)
and $-1.25\pm0.25$ at the May peak. Across the band the January value
is a flux change of only 4 per cent. The indices also inherit the
catalogue indices adopted for the flux calibrators, $-0.521$ for
J2258$-$2758 in January and $-0.713$ for J1924$-$2914 on May 26
(Section~\ref{sec:cal}), and an error in either shifts every index of
that execution by the same amount. The calibrator catalogue indices
change by 0.04 between dates, and that of the phase calibrator
J2036$-$2830 by 0.12 between January and May, so we allow an assumed
calibration uncertainty of $\pm0.1$ per epoch, equivalent to 1.1 per
cent in the relative flux scale across the band. The January peak
index is then $-0.40\pm0.18\pm0.1$. The May peak index is formally
steeper by 0.85, but we do not interpret the difference or derive
electron spectra from these indices. For the January decay we use the
same time mask as the decay fit (60\,s to the end of E3, omitting the
9--16-min rebrightening), subtract each window's own quiescent
reference, and fit the mean excess in the three windows. This gives
$\alpha=-0.96\pm0.29\pm0.1$, steeper than at the peak by
$0.56\pm0.34$ (1.6$\sigma$). The statistical error propagates the
measured noise covariance of the three windows, allowing for its
autocorrelation in time; resampling common 255-s blocks gives a smaller
$\pm0.19$. Including the rebrightening changes the index only to
$-0.89$, and the rebrightening alone gives $-0.72\pm0.44$. Using only
the tail after 5\,min, when the fast component has faded below
0.1\,mJy, gives $-0.96\pm0.42$. These are spectra of selected time
intervals, not a spectral decomposition of the two exponential
components. Negative millimetre indices are typical of
M-dwarf flares, whereas flares of the solar-type star
$\epsilon$~Eridani have a positive index \citep{Burton2022}. The 12 per cent fractional bandwidth limits
all of these indices, and simultaneous observations in two ALMA bands
would measure them far better.

The parallel hands differ by at most 1.3 per cent of Stokes~$I$ at the
January peak and by 3--4 per cent, within the noise, at the May bursts,
whereas \citet{MacGregor2020} measured $|Q/I|>0.12$ for the primary
2015 flare. Because we measure only one projection of the linear
polarisation, this is not a physical contrast, and full polarisation
observations are needed to test the mechanism.

The two flares also differ in shape. January had one dominant spike, a
separate rebrightening and a half-hour tail; May had two peaks 86\,s
apart and an 8-min tail. Could the slow January component be thermal?
Optically thin free--free emission from plasma at 10--30\,MK would need
an emission measure of (1.4--2.3)$\times10^{54}$\,cm$^{-3}$ to give the
2.5-mJy tail, and would have $\alpha\approx-0.1$. For the January decay
from 60\,s onward, excluding the rebrightening, the measured index is
$-0.96\pm0.29$ (statistical), with an assumed calibration uncertainty
of 0.1; this differs from the free--free value by 2.8$\sigma$. At 60\,s,
however, the fitted fast component still contributes about 6.4\,mJy,
against 2.4\,mJy from the slow one. After 5\,min, when the fast
component has faded, the measured index is $-0.96\pm0.42$
(statistical). With the adopted quiescent spectrum (a flat stellar
spectrum across the three windows) and relative calibration
uncertainty, treated as one standard deviation, this is about two
standard deviations steeper than optically thin free--free emission.
Together with the disc-filling brightness temperature of the tail,
(6--8)$\times10^5$\,K, this means that a thermal contribution to the
slow tail remains viable. Electrons radiating at 106\,GHz
($\nu\approx\gamma^2\nu_B$) lose energy to synchrotron emission in
$t_{\rm s}\approx7.7\times10^8/(B^2\gamma)$\,s, with $B$ in gauss,
which is about 45\,s in a 2-kG field, close to $\tau_1$. In a
single-injection model in which synchrotron cooling sets the decay
time, a time-scale of about 30\,min corresponds to a field of order
170\,G and $\gamma\approx15$; these are the characteristic values when
synchrotron cooling alone sets the 30-min time-scale, and a weaker
field would require a larger $\gamma$ at the fixed observing frequency.
Coulomb collisions shorten the electron lifetime further,
$t_{\rm c}\approx2.5\times10^{13}(\gamma/15)/n_e$\,s. Requiring
$t_{\rm c}\gtrsim1800$\,s gives
$n_e\lesssim1.4\times10^{10}(\gamma/15)$\,cm$^{-3}$ in the trapping
region; in a single-injection picture 170\,G is then an upper limit on
the field. Continued
acceleration or injection could instead sustain the emission in
stronger fields. Electrons with $\gamma\approx15$ are
ultrarelativistic, so the slow component may be closer to synchrotron
than to mildly relativistic gyrosynchrotron emission. In solar flares,
by contrast, a short non-thermal impulsive phase at millimetre
wavelengths is usually followed by an extended phase dominated by
thermal free--free emission from the hot flare plasma
\citep{Trottet2002,Luthi2004}; see \citet{Krucker2013} for a review. Successive releases of
magnetic energy into a common trapped population could explain both
the May pair of peaks and the January rebrightening. In January 73 per
cent of the recorded energy arrived after the first 90\,s; in May the
two bursts carry 59 per cent, although the 42-s scan pattern hides much
of the May decay (Section~\ref{sec:energy}).

\subsection{The quiescent millimetre excess}
\label{sec:qdisc}
On May 17 the quiet star was about 2.6 times brighter at 3\,mm than the
centimetre gyrosynchrotron spectrum of three days earlier predicts, its
3-mm flux changes by about 10 per cent between epochs, and it was
probably fainter on January 6. Because the centimetre and millimetre
data are not simultaneous and the star varies, this allows but does
not require a separate millimetre component. Spread over the stellar
disc, 0.28\,mJy at 105.6\,GHz is a brightness temperature of
(7--9)$\times10^4$\,K, about ten times the 7300-K brightness temperature
of the quiet Sun at 3\,mm \citep{White2017} and hotter than the
chromospheric layers that produce the millimetre continuum of quiet
cool stars. This points to a possible non-thermal contribution, for
example optically thick gyrosynchrotron or gyroresonance emission from
compact, strong-field active regions whose coverage changes as they
evolve. Optically thick chromospheric emission would instead give a
spectral index of $+1$ to $+2$ across the millimetre band, so
simultaneous 3- and 1-mm measurements could separate the two.
Unresolved micro-flaring, which \citet{Tristan2026b} invoke for the
12--26\,GHz quiescent emission of AU\,Mic, would produce both the
excess and its day-to-day changes. A similar mixture of chromospheric
and non-thermal emission has been proposed for the quiescent
millimetre spectrum of the young M dwarf AD\,Leo \citep{Mohan2025}.
Simultaneous centimetre and millimetre monitoring through a rotation,
with polarisation, would separate the thermal and non-thermal parts.

\subsection{Implications for debris-disc photometry}
\label{sec:implic}
The star affects single-epoch millimetre photometry of the belt in two
ways: through flares, which in this campaign raised the central flux
by factors of tens, and through quiescent changes of about 10 per cent
between epochs. Neither is removed by flagging short spikes, because
the January emission stayed elevated for 40\,min, with 35.1\,min of E3
above the threshold.

Fitting all scans of each flare execution with one constant star and
one constant belt gives the belt offset from the six-epoch quiescent
mean ($0.943\pm0.015$\,mJy), $D=100\,(B_{\rm flare}/\bar B_{\rm quiet}-1)$.
The uncertainties are in percentage points. The statistical term
combines the two fitted belt errors (for January a jackknife over ten
time blocks, as for E1). The flux-scale term is 5 per cent on each side
for January, whose calibrator differs from May--June, and an
independent 5 per cent per date for May, with E8 sharing its May 26
gain with E9. In January $D=+5.4\pm3.5$ (statistical) $\pm7.5$ (flux
scale), i.e. $+5.4\pm8.2$; in May $D=-3.7\pm4.8\pm4.4$, i.e.
$-3.7\pm6.5$. A uniform annulus gives $+12.5\pm8.7$ and $-5.1\pm6.7$.
As sensitivity tests, letting the stellar flux vary every 60\,s gives
$+6.4$ per cent in January, and shifting the field by the January
offset (Section~\ref{sec:valid}) gives $+1.9$ per cent. Comparisons
within a day remove most of the uv-coverage difference, and the flux
scale where the calibrator is shared. E3 against the mean of E2 and E4
gives $+11\pm6$ per cent ($+6\pm6$ with the field shifted), including
the independent 5-per-cent scale of E2, which was calibrated on
J1617$-$5848; E2 and E4 themselves differ by $8\pm7$ per cent. E8
against E9, which share their calibrator and date, gives $-7\pm6$ per
cent.
The execution-averaged belt flux therefore shows no significant
change during the flares, although these fits do not constrain changes
on the time-scale of a single spike; no comparison differs from zero by much more than
$2\sigma$. The January stellar excess persisted after the brightest
integrations were removed. Fitting a separate stellar amplitude in each
execution gave belt fluxes consistent with the six-epoch quiescent
mean.
Simultaneous ALMA and optical, ultraviolet and X-ray monitoring would
tie the millimetre flares to the thermal flare energy; AU\,Mic flares
often at optical wavelengths \citep{Gilbert2022,Tristan2023}, so such
campaigns are practical. Millimetre flares have also been seen from the
debris-disc host $\epsilon$~Eridani \citep{Burton2022}, so the same
caution applies to belt photometry of other active hosts.

\subsection{Relevance to the planets}
\label{sec:planets}
AU\,Mic\,b orbits at about 0.065\,au, 17--19 stellar radii
\citep{Plavchan2020}. If the millimetre emission is gyrosynchrotron
radiation, the January decay is consistent with energetic electrons
being accelerated or confined for tens of minutes. These observations
do not show whether any particles escaped or reached the planets.

\section{Conclusions}
\label{sec:concl}
Twelve ALMA Band~3 executions (7.8\,h on source, 2025 January--June)
answer our three questions and give a first census of 3-mm flaring.
\begin{enumerate}
\item \textbf{The quiet star.} Separated from the belt in the
visibilities, AU\,Mic has a mean May--June 3-mm flux density of
$0.28\pm0.02$\,mJy, about 20 times its photosphere. Two estimators
give intrinsic epoch-to-epoch rms amplitudes of 8 and 11 per cent of
the mean over six epochs; a test that does not depend on the flux
scale rejects a constant star at $p=0.001$--0.003. On May 17 it was
about 2.6 times the 106-GHz extrapolation of a VLA spectrum taken three
days earlier, and its disc-averaged brightness temperature,
(7--9)$\times10^4$\,K, leaves room for a non-thermal contribution. On
January 6 it was probably fainter, $0.16\pm0.03\pm0.02$\,mJy
(2.4--2.6$\sigma$ below May--June). The belt, best described by a brightness rising outward as
$r^{2.3}$ (adopted index 2.32), has $0.94\pm0.05$\,mJy and a 1.3--3\,mm spectral index of
$2.53\pm0.15$.
\item \textbf{The flares.} On January 19 the brightest recorded
integration, the first of E3 after a calibration gap, was
$29.2\pm0.4\pm1.5$\,mJy ($3.3\times10^{15}$\,erg\,s$^{-1}$\,Hz$^{-1}$);
the rise was unobserved, and the true maximum may have been higher.
The flux fell by a factor of 4.6 in 90\,s and then decayed with a
fitted slow e-folding time of 30--38\,min, depending on the fit window
and quiescent level. The recorded Band~3 excess energy was
(2.8--3.2)$\times10^{27}$\,erg, with a further 5 per cent flux-scale
uncertainty. On May 26 two peaks 86\,s apart reached 10.1 and
10.6\,mJy (total flux),
followed by a decay with a fitted slow e-folding time of 8.1\,min and a
recorded energy of $6.8\times10^{26}$\,erg. Both events are consistent
with compact emission from AU\,Mic. With the rebrightening excluded, the
January decay from 60\,s onward has an in-band index of
$-0.96\pm0.29$ (statistical), but this interval still includes the
fast component. After five minutes the measured index is
$-0.96\pm0.42$ (statistical). With the adopted quiescent spectrum and
relative calibration uncertainty, this is about two standard deviations
steeper than optically thin free--free emission. A thermal contribution
to the slow tail remains viable. Equating the approximately
30-min slow decay to the synchrotron loss time of electrons radiating
at 106\,GHz gives a characteristic field of about 170\,G in a
single-injection model. Continued injection would permit stronger
fields. A tail as bright as
the January one, scaled to 220\,GHz, was absent 3--5.4\,min after the
peak of the 2015 flare, which we recover.
\item \textbf{Occurrence.} Two events with recorded integrations above
10\,mJy were seen in 7.81\,h on source. The star was above our threshold for 15.8 per
cent of the sampled exposure, and for 3.3 per cent outside January 19.
\item \textbf{The belt.} Flagging the brightest integrations leaves a
prolonged stellar excess. With a separate stellar amplitude fitted in
each execution, the inferred belt flux differed from its quiescent mean
by $+5.4\pm8.2$ per cent in January
and $-3.7\pm6.5$ per cent in May, including flux-scale uncertainty
($+12.5\pm8.7$ and $-5.1\pm6.7$ per cent for a uniform belt); neither
difference is significant. These are execution-averaged values and do
not constrain belt changes on the time-scale of a single spike.
\end{enumerate}

\section*{Acknowledgements}
\sloppy
This paper makes use of the following ALMA data:
ADS/JAO.ALMA\#2024.1.01095.S, \#2016.1.00878.S, \#2012.1.00198.S
and \#2017.1.01583.S. ALMA is a partnership of ESO
(representing its member states), NSF (USA) and NINS (Japan), together
with NRC (Canada), NSTC and ASIAA (Taiwan), and KASI (Republic of
Korea), in cooperation with the Republic of Chile. The Joint ALMA
Observatory is operated by ESO, AUI/NRAO and NAOJ. This work used
\textsc{casa} \citep{CASA2022}, NumPy \citep{Harris2020}, SciPy
\citep{Virtanen2020} and Matplotlib \citep{Hunter2007}.

\textit{AI use and validation.} The data reduction, analysis, text and
figures were produced by the author with assistance from the Python
framework ASTRA \citep{White2026}. This AI-assisted system for
astrophysical analysis was carefully validated by reproduction of
published comparison-star fluxes, recovery of the known 2015 flare
(detected at high significance, although its published peak amplitude
is not reproduced) and synthetic-flare injection tests. The author
designed and verified the analysis, conclusions and text, is
responsible for any errors, and declares no conflict of interest.

\section*{Data availability}
The ALMA data are public in the ALMA Science Archive. Machine-readable
tables accompany the arXiv submission as ancillary files:
\texttt{table\_s1\_perscan.csv} (per-scan photometry of all 1169
Band~3 scans, including calibrator and atmosphere scans, with intent, reference level, excess and threshold
class; the 457 science scans analysed are those with
\texttt{scan\_class}\,=\,\texttt{science}), \texttt{table\_s2\_b9\_perscan.csv} and
\texttt{table\_s3\_b6\_perscan.csv} (the archival controls),
\texttt{table\_s4\_stellar.csv} (the star and belt fits),
\texttt{table\_s5\_spw.csv} (the spectral-window excess fluxes behind
the in-band indices; \texttt{table\_s5b\_spw\_masked.csv} for the
masked January decay), the twelve
per-integration light curves \texttt{dumps\_<execution>.csv}, the
analysis scripts (\texttt{scripts/}) and the fitted flare positions
(\texttt{table\_s6\_positions.csv}). The visibility extracts (2\,GB) and
per-integration stores on which the scripts operate (the extracts hold
the real part of the phase-centred Stokes~$I$ visibilities), with the
intermediate result files, are deposited in Zenodo
(\href{https://doi.org/\zenododoi}{doi:\zenododoi}).

\bibliographystyle{mnras}
\bibliography{refs}

\begin{thebibliography}{}
\makeatletter
\relax
\def\mn@urlcharsother{\let\do\@makeother \do\$\do\&\do\#\do\^\do\_\do\%\do\~}
\def\mn@doi{\begingroup\mn@urlcharsother \@ifnextchar [ {\mn@doi@}
  {\mn@doi@[]}}
\def\mn@doi@[#1]#2{\def\@tempa{#1}\ifx\@tempa\@empty \href
  {http://dx.doi.org/#2} {doi:#2}\else \href {http://dx.doi.org/#2} {#1}\fi
  \endgroup}
\def\mn@eprint#1#2{\mn@eprint@#1:#2::\@nil}
\def\mn@eprint@arXiv#1{\href {http://arxiv.org/abs/#1} {{\tt arXiv:#1}}}
\def\mn@eprint@dblp#1{\href {http://dblp.uni-trier.de/rec/bibtex/#1.xml}
  {dblp:#1}}
\def\mn@eprint@#1:#2:#3:#4\@nil{\def\@tempa {#1}\def\@tempb {#2}\def\@tempc
  {#3}\ifx \@tempc \@empty \let \@tempc \@tempb \let \@tempb \@tempa \fi \ifx
  \@tempb \@empty \def\@tempb {arXiv}\fi \@ifundefined
  {mn@eprint@\@tempb}{\@tempb:\@tempc}{\expandafter \expandafter \csname
  mn@eprint@\@tempb\endcsname \expandafter{\@tempc}}}

\bibitem[\protect\citeauthoryear{{Bastian}, {Benz}  \& {Gary}}{{Bastian}
  et~al.}{1998}]{Bastian1998}
{Bastian} T.~S.,  {Benz} A.~O.,   {Gary} D.~E.,  1998, \mn@doi [\araa]
  {10.1146/annurev.astro.36.1.131}, 36, 131

\bibitem[\protect\citeauthoryear{{Benz} \& {G{\"u}del}}{{Benz} \&
  {G{\"u}del}}{2010}]{Benz2010}
{Benz} A.~O.,  {G{\"u}del} M.,  2010, \mn@doi [\araa]
  {10.1146/annurev-astro-082708-101757}, 48, 241

\bibitem[\protect\citeauthoryear{{Biermann}, {Li}, {Naess}  et~al.}{{Biermann}
  et~al.}{2025}]{Biermann2025}
{Biermann} E.,  {Li} Y.,  {Naess} S.,   et~al., 2025, \mn@doi [\apj]
  {10.3847/1538-4357/adce70}, 986, 17

\bibitem[\protect\citeauthoryear{{Boldog} et~al.}{{Boldog}
  et~al.}{2025}]{Boldog2025}
{Boldog} {\AA}.,  et~al., 2025, \mn@doi [\aap] {10.1051/0004-6361/202452699},
  694, A137

\bibitem[\protect\citeauthoryear{{Bower}, {Plambeck}, {Bolatto}, {McCrady},
  {Graham}, {de Pater}, {Liu}  \& {Baganoff}}{{Bower} et~al.}{2003}]{Bower2003}
{Bower} G.~C.,  {Plambeck} R.~L.,  {Bolatto} A.,  {McCrady} N.,  {Graham}
  J.~R.,  {de Pater} I.,  {Liu} M.~C.,   {Baganoff} F.~K.,  2003, \mn@doi
  [\apj] {10.1086/379101}, 598, 1140

\bibitem[\protect\citeauthoryear{{Burton}, {MacGregor}  \& {Osten}}{{Burton}
  et~al.}{2022}]{Burton2022}
{Burton} K.,  {MacGregor} M.~A.,   {Osten} R.~A.,  2022, \mn@doi [\apjl]
  {10.3847/2041-8213/ac9973}, 939, L6

\bibitem[\protect\citeauthoryear{{Burton} et~al.,}{{Burton}
  et~al.}{2025}]{Burton2025}
{Burton} K.,  et~al., 2025, \mn@doi [\apj] {10.3847/1538-4357/ada5f2}, 982, 43

\bibitem[\protect\citeauthoryear{{CASA Team} et~al.}{{CASA Team}
  et~al.}{2022}]{CASA2022}
{CASA Team} et~al., 2022, \mn@doi [\pasp] {10.1088/1538-3873/ac9642}, 134,
  114501

\bibitem[\protect\citeauthoryear{{Cronin-Coltsmann} et~al.,}{{Cronin-Coltsmann}
  et~al.}{2022}]{Cronin2022}
{Cronin-Coltsmann} P.~F.,  et~al., 2022, \mn@doi [\mnras]
  {10.1093/mnras/stac536}, 512, 4752

\bibitem[\protect\citeauthoryear{{Cronin-Coltsmann}, {Kennedy}, {Kral},
  {Lestrade}, {Marino}, {Matr{\`a}}  \& {Wyatt}}{{Cronin-Coltsmann}
  et~al.}{2023}]{Cronin2023}
{Cronin-Coltsmann} P.~F.,  {Kennedy} G.~M.,  {Kral} Q.,  {Lestrade} J.-F.,
  {Marino} S.,  {Matr{\`a}} L.,   {Wyatt} M.~C.,  2023, \mn@doi [\mnras]
  {10.1093/mnras/stad3083}, 526, 5401

\bibitem[\protect\citeauthoryear{{Daley}, {Hughes}, {Carter}  et~al.}{{Daley}
  et~al.}{2019}]{Daley2019}
{Daley} C.,  {Hughes} A.~M.,  {Carter} E.~S.,   et~al., 2019, \mn@doi [\apj]
  {10.3847/1538-4357/ab1074}, 875, 87

\bibitem[\protect\citeauthoryear{{Donati} et~al.}{{Donati}
  et~al.}{2023}]{Donati2023}
{Donati} J.-F.,  et~al., 2023, \mn@doi [\mnras] {10.1093/mnras/stad1193}, 525,
  455

\bibitem[\protect\citeauthoryear{{Donati} et~al.}{{Donati}
  et~al.}{2025}]{Donati2025}
{Donati} J.-F.,  et~al., 2025, \mn@doi [\aap] {10.1051/0004-6361/202555428},
  700, A122

\bibitem[\protect\citeauthoryear{{Dulk}}{{Dulk}}{1985}]{Dulk1985}
{Dulk} G.~A.,  1985, \mn@doi [\araa] {10.1146/annurev.aa.23.090185.001125}, 23,
  169

\bibitem[\protect\citeauthoryear{{Gaia Collaboration}, {Vallenari}
  et~al.}{{Gaia Collaboration} et~al.}{2023}]{Gaia2023}
{Gaia Collaboration} {Vallenari} A.,   et~al., 2023, \mn@doi [\aap]
  {10.1051/0004-6361/202243940}, 674, A1

\bibitem[\protect\citeauthoryear{{Gilbert} et~al.}{{Gilbert}
  et~al.}{2022}]{Gilbert2022}
{Gilbert} E.~A.,  et~al., 2022, \mn@doi [\aj] {10.3847/1538-3881/ac23ca}, 163,
  147

\bibitem[\protect\citeauthoryear{{G{\"u}del}}{{G{\"u}del}}{2002}]{Gudel2002}
{G{\"u}del} M.,  2002, \mn@doi [\araa]
  {10.1146/annurev.astro.40.060401.093806}, 40, 217

\bibitem[\protect\citeauthoryear{{Guns}, {Foster}, {Daley}  et~al.}{{Guns}
  et~al.}{2021}]{Guns2021}
{Guns} S.,  {Foster} A.,  {Daley} C.,   et~al., 2021, \mn@doi [\apj]
  {10.3847/1538-4357/ac06a3}, 916, 98

\bibitem[\protect\citeauthoryear{{Harris}, {Millman}, {van der Walt}
  et~al.}{{Harris} et~al.}{2020}]{Harris2020}
{Harris} C.~R.,  {Millman} K.~J.,  {van der Walt} S.~J.,   et~al., 2020,
  \mn@doi [Nature] {10.1038/s41586-020-2649-2}, 585, 357

\bibitem[\protect\citeauthoryear{{Howard}, {MacGregor}, {Osten}
  et~al.}{{Howard} et~al.}{2022}]{Howard2022}
{Howard} W.~S.,  {MacGregor} M.~A.,  {Osten} R.,   et~al., 2022, \mn@doi [\apj]
  {10.3847/1538-4357/ac9134}, 938, 103

\bibitem[\protect\citeauthoryear{{Hunter}}{{Hunter}}{2007}]{Hunter2007}
{Hunter} J.~D.,  2007, \mn@doi [Computing in Science and Engineering]
  {10.1109/MCSE.2007.55}, 9, 90

\bibitem[\protect\citeauthoryear{{Kalas}, {Liu}  \& {Matthews}}{{Kalas}
  et~al.}{2004}]{Kalas2004}
{Kalas} P.,  {Liu} M.~C.,   {Matthews} B.~C.,  2004, \mn@doi [Science]
  {10.1126/science.1093420}, 303, 1990

\bibitem[\protect\citeauthoryear{{Kochukhov} \& {Reiners}}{{Kochukhov} \&
  {Reiners}}{2020}]{Kochukhov2020}
{Kochukhov} O.,  {Reiners} A.,  2020, \mn@doi [\apj]
  {10.3847/1538-4357/abb2a2}, 902, 43

\bibitem[\protect\citeauthoryear{{Krucker}, {Gim{\'e}nez de Castro}, {Hudson}
  et~al.}{{Krucker} et~al.}{2013}]{Krucker2013}
{Krucker} S.,  {Gim{\'e}nez de Castro} C.~G.,  {Hudson} H.~S.,   et~al., 2013,
  \mn@doi [\aapr] {10.1007/s00159-013-0058-3}, 21, 58

\bibitem[\protect\citeauthoryear{{L{\"u}thi}, {Magun}  \& {Miller}}{{L{\"u}thi}
  et~al.}{2004}]{Luthi2004}
{L{\"u}thi} T.,  {Magun} A.,   {Miller} M.,  2004, \mn@doi [\aap]
  {10.1051/0004-6361:20034624}, 415, 1123

\bibitem[\protect\citeauthoryear{{MacGregor} et~al.}{{MacGregor}
  et~al.}{2013}]{MacGregor2013}
{MacGregor} M.~A.,  et~al., 2013, \mn@doi [\apjl]
  {10.1088/2041-8205/762/2/L21}, 762, L21

\bibitem[\protect\citeauthoryear{{MacGregor} et~al.,}{{MacGregor}
  et~al.}{2016}]{MacGregor2016}
{MacGregor} M.~A.,  et~al., 2016, \mn@doi [\apj] {10.3847/0004-637X/823/2/79},
  823, 79

\bibitem[\protect\citeauthoryear{{MacGregor}, {Weinberger}, {Wilner},
  {Kowalski}  \& {Cranmer}}{{MacGregor} et~al.}{2018}]{MacGregor2018}
{MacGregor} M.~A.,  {Weinberger} A.~J.,  {Wilner} D.~J.,  {Kowalski} A.~F.,
  {Cranmer} S.~R.,  2018, \mn@doi [\apjl] {10.3847/2041-8213/aaad6b}, 855, L2

\bibitem[\protect\citeauthoryear{{MacGregor} et~al.}{{MacGregor}
  et~al.}{2020}]{MacGregor2020}
{MacGregor} M.~A.,  et~al., 2020, \mn@doi [\apj] {10.3847/1538-4357/ab711d},
  891, 80

\bibitem[\protect\citeauthoryear{{MacGregor} et~al.}{{MacGregor}
  et~al.}{2021}]{MacGregor2021}
{MacGregor} M.~A.,  et~al., 2021, \mn@doi [\apjl] {10.3847/2041-8213/abf14c},
  911, L25

\bibitem[\protect\citeauthoryear{{Mamajek} \& {Bell}}{{Mamajek} \&
  {Bell}}{2014}]{Mamajek2014}
{Mamajek} E.~E.,  {Bell} C.~P.~M.,  2014, \mn@doi [\mnras]
  {10.1093/mnras/stu1894}, 445, 2169

\bibitem[\protect\citeauthoryear{{Marino}, {Matr{\`a}}, {Hughes}
  et~al.}{{Marino} et~al.}{2026}]{Marino2026}
{Marino} S.,  {Matr{\`a}} L.,  {Hughes} A.~M.,   et~al., 2026, \mn@doi [\aap]
  {10.1051/0004-6361/202556489}, 705, A195

\bibitem[\protect\citeauthoryear{{Martioli}, {H{\'e}brard}, {Correia}, {Laskar}
   \& {Lecavelier des Etangs}}{{Martioli} et~al.}{2021}]{Martioli2021}
{Martioli} E.,  {H{\'e}brard} G.,  {Correia} A.~C.~M.,  {Laskar} J.,
  {Lecavelier des Etangs} A.,  2021, \mn@doi [\aap]
  {10.1051/0004-6361/202040235}, 649, A177

\bibitem[\protect\citeauthoryear{{Massi}, {Forbrich}, {Menten},
  {Torricelli-Ciamponi}, {Neidh{\"o}fer}, {Leurini}  \& {Bertoldi}}{{Massi}
  et~al.}{2006}]{Massi2006}
{Massi} M.,  {Forbrich} J.,  {Menten} K.~M.,  {Torricelli-Ciamponi} G.,
  {Neidh{\"o}fer} J.,  {Leurini} S.,   {Bertoldi} F.,  2006, \mn@doi [\aap]
  {10.1051/0004-6361:20053535}, 453, 959

\bibitem[\protect\citeauthoryear{{Matthews}, {Kennedy}, {Sibthorpe}, {Holland},
  {Booth}, {Kalas}  et~al.}{{Matthews} et~al.}{2015}]{Matthews2015}
{Matthews} B.~C.,  {Kennedy} G.,  {Sibthorpe} B.,  {Holland} W.,  {Booth} M.,
  {Kalas} P.,   et~al., 2015, \mn@doi [\apj] {10.1088/0004-637X/811/2/100},
  811, 100

\bibitem[\protect\citeauthoryear{{Mohan}, {Hauschildt}, {Fuhrmeister}
  et~al.}{{Mohan} et~al.}{2025}]{Mohan2025}
{Mohan} A.,  {Hauschildt} P.~H.,  {Fuhrmeister} B.,   et~al., 2025, \mn@doi
  [\apj] {10.3847/1538-4357/ade9bf}, 989, 20

\bibitem[\protect\citeauthoryear{{Naess}, {Battaglia}, {Bond}  et~al.}{{Naess}
  et~al.}{2021}]{Naess2021}
{Naess} S.,  {Battaglia} N.,  {Bond} J.~R.,   et~al., 2021, \mn@doi [\apj]
  {10.3847/1538-4357/abfe6d}, 915, 14

\bibitem[\protect\citeauthoryear{{Plavchan}, {Barclay}, {Gagn{\'e}}
  et~al.}{{Plavchan} et~al.}{2020}]{Plavchan2020}
{Plavchan} P.,  {Barclay} T.,  {Gagn{\'e}} J.,   et~al., 2020, \mn@doi [\nat]
  {10.1038/s41586-020-2400-z}, 582, 497

\bibitem[\protect\citeauthoryear{{Ricker}, {Winn}, {Vanderspek}
  et~al.}{{Ricker} et~al.}{2015}]{Ricker2015}
{Ricker} G.~R.,  {Winn} J.~N.,  {Vanderspek} R.,   et~al., 2015, \mn@doi [J.
  Astron. Telesc. Instrum. Syst.] {10.1117/1.JATIS.1.1.014003}, 1, 014003

\bibitem[\protect\citeauthoryear{{Salter}, {Hogerheijde}  \& {Blake}}{{Salter}
  et~al.}{2008}]{Salter2008}
{Salter} D.~M.,  {Hogerheijde} M.~R.,   {Blake} G.~A.,  2008, \mn@doi [\aap]
  {10.1051/0004-6361:200810807}, 492, L21

\bibitem[\protect\citeauthoryear{{Salter}, {K{\'o}sp{\'a}l}, {Getman},
  {Hogerheijde}, {van Kempen}, {Carpenter}, {Blake}  \& {Wilner}}{{Salter}
  et~al.}{2010}]{Salter2010}
{Salter} D.~M.,  {K{\'o}sp{\'a}l} {\'A}.,  {Getman} K.~V.,  {Hogerheijde}
  M.~R.,  {van Kempen} T.~A.,  {Carpenter} J.~M.,  {Blake} G.~A.,   {Wilner}
  D.,  2010, \mn@doi [\aap] {10.1051/0004-6361/201015197}, 521, A32

\bibitem[\protect\citeauthoryear{{Szab{\'o}} et~al.}{{Szab{\'o}}
  et~al.}{2021}]{Szabo2021}
{Szab{\'o}} G.~M.,  et~al., 2021, \mn@doi [\aap] {10.1051/0004-6361/202140345},
  654, A159

\bibitem[\protect\citeauthoryear{{Tandoi}, {Guns}, {Foster}  et~al.}{{Tandoi}
  et~al.}{2024}]{Tandoi2024}
{Tandoi} C.,  {Guns} S.,  {Foster} A.,   et~al., 2024, \mn@doi [\apj]
  {10.3847/1538-4357/ad58db}, 972, 6

\bibitem[\protect\citeauthoryear{{Tristan}, {Notsu}, {Kowalski}, {Brown}
  et~al.}{{Tristan} et~al.}{2023}]{Tristan2023}
{Tristan} I.~I.,  {Notsu} Y.,  {Kowalski} A.~F.,  {Brown} A.,   et~al., 2023,
  \mn@doi [\apj] {10.3847/1538-4357/acc94f}, 951, 33

\bibitem[\protect\citeauthoryear{{Tristan}, {Osten}  et~al.}{{Tristan}
  et~al.}{2025}]{Tristan2025}
{Tristan} I.~I.,  {Osten} R.~A.,   et~al., 2025, \mn@doi [\apj]
  {10.3847/1538-4357/adc565}, 986, 53

\bibitem[\protect\citeauthoryear{{Tristan} et~al.}{{Tristan}
  et~al.}{2026a}]{Tristan2026}
{Tristan} I.~I.,  et~al., 2026a, \mn@doi [\apj] {10.3847/1538-4357/ae2857},
  997, 168

\bibitem[\protect\citeauthoryear{{Tristan}, {Osten}, {Notsu}, {Feinstein}  \&
  {Kowalski}}{{Tristan} et~al.}{2026b}]{Tristan2026b}
{Tristan} I.~I.,  {Osten} R.~A.,  {Notsu} Y.,  {Feinstein} A.~D.,   {Kowalski}
  A.~F.,  2026b, \mn@doi [\apj] {10.3847/1538-4357/ae9d50}, 1009, 212

\bibitem[\protect\citeauthoryear{{Trottet}, {Raulin}, {Kaufmann}, {Siarkowski},
  {Klein}  \& {Gary}}{{Trottet} et~al.}{2002}]{Trottet2002}
{Trottet} G.,  {Raulin} J.-P.,  {Kaufmann} P.,  {Siarkowski} M.,  {Klein}
  K.-L.,   {Gary} D.~E.,  2002, \mn@doi [\aap] {10.1051/0004-6361:20011556},
  381, 694

\bibitem[\protect\citeauthoryear{{Virtanen}, {Gommers}, {Oliphant}
  et~al.}{{Virtanen} et~al.}{2020}]{Virtanen2020}
{Virtanen} P.,  {Gommers} R.,  {Oliphant} T.~E.,   et~al., 2020, \mn@doi
  [Nature Methods] {10.1038/s41592-019-0686-2}, 17, 261

\bibitem[\protect\citeauthoryear{{Vizgan}, {Hughes}, {Carter}, {Flaherty},
  {Pan}, {Chiang}  et~al.}{{Vizgan} et~al.}{2022}]{Vizgan2022}
{Vizgan} D.,  {Hughes} A.~M.,  {Carter} E.~S.,  {Flaherty} K.~M.,  {Pan} M.,
  {Chiang} E.,   et~al., 2022, \mn@doi [\apj] {10.3847/1538-4357/ac80b8}, 935,
  131

\bibitem[\protect\citeauthoryear{{White}}{{White}}{2026}]{White2026}
{White} G.~J.,  2026, \mn@doi [RAS Techniques and Instruments]
  {10.1093/rasti/rzag088}, 5, rzag088

\bibitem[\protect\citeauthoryear{{White}, {Iwai}, {Phillips}  et~al.}{{White}
  et~al.}{2017}]{White2017}
{White} S.~M.,  {Iwai} K.,  {Phillips} N.~M.,   et~al., 2017, \mn@doi
  [\solphys] {10.1007/s11207-017-1123-2}, 292, 88

\bibitem[\protect\citeauthoryear{{Wittrock} et~al.}{{Wittrock}
  et~al.}{2022}]{Wittrock2022}
{Wittrock} J.~M.,  et~al., 2022, \mn@doi [\aj] {10.3847/1538-3881/ac68e5}, 164,
  27

\bibitem[\protect\citeauthoryear{{Wootten} \& {Thompson}}{{Wootten} \&
  {Thompson}}{2009}]{Wootten2009}
{Wootten} A.,  {Thompson} A.~R.,  2009, \mn@doi [Proceedings of the IEEE]
  {10.1109/jproc.2009.2020572}, 97, 1463

\makeatother
\end{thebibliography}

\end{document}